**Two-Path Solid-State Interferometry Using Ultra-Subwavelength 2D Plasmonic Waves**

Kitty Y. M. Yeung[1], Hosang Yoon[1], William Andress[1], Ken West[2], Loren Pfeiffer[2] and Donhee Ham[1,a)]

[1]School of Engineering and Applied Sciences, Harvard University, 33 Oxford St, Cambridge, MA 02138, USA.

[2]Department of Electrical Engineering, Princeton University, Princeton, New Jersey 08544, USA.

We report an on-chip solid-state Mach-Zehnder interferometer operating on two-dimensional (2D) plasmonic waves at microwave frequencies. Two plasmonic paths are defined with GaAs/AlGaAs 2D electron gas 80 nm below a metallic gate. The gated 2D plasmonic waves achieve a velocity of $\sim c/300$ ($c$: free-space light speed). Due to this ultra-subwavelength confinement, the resolution of the 2D plasmonic interferometer is two orders of magnitude higher than that of its electromagnetic counterpart at a given frequency. This GHz proof-of-concept at cryogenic temperatures can be scaled to the THz~IR range for room temperature operation, while maintaining the benefits of the ultra-subwavelength confinement.

Plasmas appear in various forms in nature, with the collective electron density waves, or plasmonic waves, serving as a salient dynamic feature. Solid-state plasmas consisting of mobile electrons in metals and semiconductors are especially interesting, from the point of view that fabrication technologies available for solid-state materials allow us to design the boundaries and interfaces of the plasma media in order to engineer the plasmonic waves. In particular, surface plasmons on three-dimensional (3D) bulk metals have been an active subject of research in photonics. One of these efforts with surface plasmons concerns developing interferometers[1-4]. A prominent advantage of surface plasmonic interferometers is their high resolution. Surface plasmons can achieve a velocity as low as $\sim c/10$ with a proportionally reduced wavelength[5].

---

a) Author to whom correspondence should be addressed. Electronic mail: donhee@seas.harvard.edu

This subwavelength confinement makes surface plasmon interferometers more sensitive to the path length difference and the surrounding dielectric media, leading to the higher resolution in measurements of physical quantities such as length and dielectric constant.

While surface plasmons on 3D bulk metals typically appear in the optics regime, plasmonic waves in 2D conductors, such as GaAs/AlGaAs 2D electron gas (2DEG) and graphene, can appear in the GHz~THz frequencies[6-12]. These 2D plasmons exhibit a greater subwavelength confinement with velocities far below[10] ~$c$/100. Harnessing this ultra-subwavelength confinement, we develop a high-resolution 2D plasmonic interferometer at microwave frequencies in the electronics regime, drawing a parallel line of development to surface plasmonic interferometers in photonics.

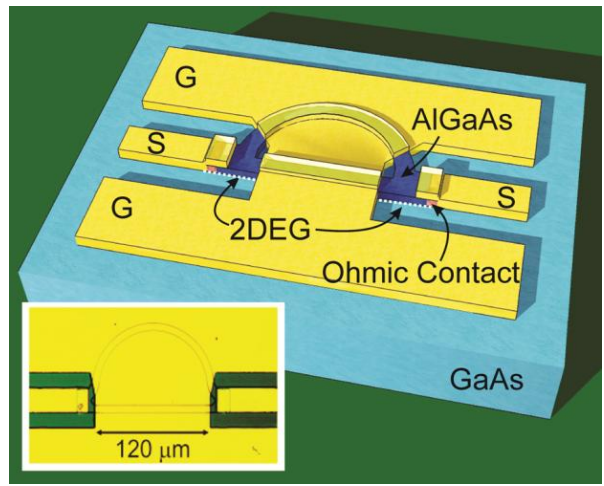

**Fig. 1:** Diagram of the 2D plasmonic Mach-Zehnder interferometer (not drawn to scale). The inset shows an optical image of an actual implementation. 2DEG forms between AlGaAs and GaAs layers.

For proof of concept, we use GaAs/AlGaAs 2DEG as a 2D plasmonic medium and construct a Mach-Zehnder interferometer, whose two paths are defined by mesa-etching the 2DEG [Fig. 1; APPENDIX A]. The curved 2DEG path has a length of $l_1$ ~ 191 μm and the linear path has a length of $l_2$ ~120 μm while both have the same width of $w$ ~ 8 μm. The left metallic coplanar waveguide (CPW)[13] consisting of a signal line ('S') and two ground lines ('G') guides an electromagnetic wave onto the left end of the interferometer; the CPW's signal line makes an Ni/Au/Ge ohmic contact to the 2DEG. The excited plasmonic wave splits into two plasmonic waves along the two 2DEG paths, which superpose at the right end of the interferometer. By this junction, the two waves develop a frequency-dependent phase difference due to the path length difference, thus, their superposition exhibits an interference pattern with frequency. This superposed wave excites an electromagnetic wave onto the right CPW, which we measure to study the interference. The characteristic impedance, $Z_0$, of both CPWs is designed to be 50 Ω.

A metallic top gate placed 80 nm above both 2DEG paths serves as an *ac* plasmonic ground [Fig. 1]; each 2DEG signal path with the gate as the *ac* ground then forms a plasmonic transmission line[10]. The plasmonic ground is merged with the CPWs' ground lines. This ground sharing ensures seamless continuation of the purely plasmonic transmission lines from the purely electromagnetic CPWs[10]; with no such ground sharing, the gate itself would act as an electromagnetic path, instead of plasmonic ground. The gate also serves to further slow the 2D plasmonic velocity (which is already small due to the reduced dimensionality) by shortening the Coulomb interaction range within 2DEG[10, 14]. This further enhances subwavelength confinement and interferometer resolution. The gated 2D plasmonic velocity[14], $v_p = \sqrt{(ne^2 d)/(m^* \kappa \varepsilon_0)}$, is ~ $c$/300 in our case as seen shortly (*n*: conduction electron density per unit area; $m^*$: effective electron mass; *d*: 2DEG-gate distance; *κ*: dielectric constant of the AlGaAs medium between the 2DEG and gate). Yet another role of the gate is to provide a *dc* bias to tune the electron density, hence $v_p$. In actuality, we apply to the CPWs' signal lines a *dc* bias with reference to the ground lines, so that the gate is effectively biased in the opposite polarity relative to the 2DEG.

We perform microwave scattering experiments [APPENDIX B] in the dark up to 50 GHz with a vector network analyzer to obtain frequency-domain interference patterns. Temperature is at 4.2 K so that plasmonic dynamics is not masked by electron scatterings. The raw *s*-parameters contain not only the interferometer effect but the effect of direct coupling between the left and right on-chip CPWs bypassing the interferometer. This parasitic coupling is separately measured and de-embedded[10, 11] [APPENDIX B], yielding *s*-parameters containing only the interferometer effect. The *s*-parameters discussed from here on are such de-embedded ones, unless otherwise stated. Theoretically, the transmission $s_{21}$ of the interferometer can be approximated as the following, where we safely consider only the 1st-order signal pathways, and for simplicity neglect the ohmic contacts and the short ungated 2DEG regions [Fig. 1; APPENDIX E]:

$$s_{21} \approx \frac{4Z_0 Z}{(Z+2Z_0)^2}\left[e^{-\alpha l_1}e^{-i\beta l_1} + e^{-\alpha l_2}e^{-i\beta l_2}\right] = \frac{4Z_0 Z}{(Z+2Z_0)^2} A_1 e^{-i\beta l_1}\left[1 + \frac{A_2}{A_1}e^{-i\Delta\phi}\right]. \quad (1)$$

$Z$ is the identical characteristic impedance of the either plasmonic line [APPENDIX D]; $\beta \equiv \omega/v_p$ and $\alpha$ are the wavenumber and attenuation constant (the latter due to electron scatterings in 2DEG) of the plasmonic lines; $A_1 \equiv e^{-\alpha l_1}$ and $A_2 \equiv e^{-\alpha l_2}$; and $\Delta\phi \equiv \beta(l_1-l_2) = \omega(l_1-l_2)/v_p$ is the phase difference between the two plasmonic waves at the right junction of the interferometer. The $s_{21}$ is proportional to the superposition of the two plasmonic waves and will exhibit an interference pattern with frequency. The factor $4Z_0Z/(Z+2Z_0)^2 \equiv F$ is complex in general, as $Z$ is generally complex due to loss caused by electron scatterings; however, $F$ exhibits a practically constant magnitude and a negligible phase over the measurement frequency range [APPENDIX D].

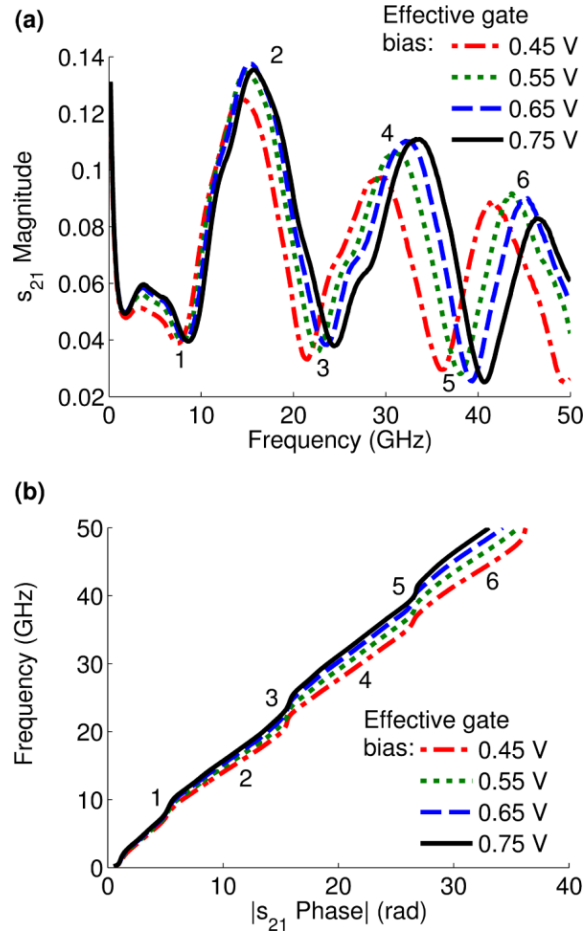

**Fig. 2:** (a) $s_{21}$ magnitude; (b) absolute value of $s_{21}$ phase at effective gate bias 0.45 V (red, semi-dashed), 0.55 V (green, dotted), 0.65 V (blue, dashed) and 0.75 V (black, solid). Temperature: 4.2 K. In (b), $y$-axis is frequency and $x$-axis is $s_{21}$ phase, following the convention of showing dispersion.

Figure 2(a) shows the magnitude of the measured $s_{21}$ vs. frequency at various effective gate biases. At each bias the $s_{21}$ magnitude shows the anticipated interference pattern with destructive dips marked as 1, 3 and 5, and constructive peaks as 2, 4 and 6. The reduction of the peak and dip magnitudes with frequency, which is not predicted by Eq. (1), stems from the frequency dependent behaviors of the ohmic contacts and the ungated 2DEG regions[10]. According to Eq. (1), the destructive dips (constructive peaks) with amplitudes proportional to $1 - (+) A_2/A_1$ occur at frequencies that make $\Delta\phi \equiv \omega(l_1-l_2)/v_p$ odd (even) integer multiples of $\pi$. The increasing gate bias increases $v_p$, leading to the shift of the interference pattern to the right [Fig. 2(a)]; for a larger $v_p$, higher frequencies are required to produce the same $\Delta\phi$. From these considerations in connection with Eq. (1), we can extract, at a given bias, $v_p$ at six different frequencies corresponding to the three dips and three peaks of the $s_{21}$ magnitude. Figure 3 shows the range (red bar) and median (red square) of such six extracted $v_p$ values at each bias, where the median is the average of the two middle $v_p$ values. While $v_p$ must be constant at a given bias, the extracted $v_p$ shows the frequency variation (bar). This is due to the frequency-dependent behaviors of the ohmic contacts and ungated 2DEG regions[10], which can shift the dip and peak positions from the ideal positions predicted by Eq. (1). While the actual $v_p$ should monotonically increase with gate bias, the extracted $v_p$ shows deviation from this pattern at biases below 0.4 V [Fig. 3]. We suspect that this abnormality is also caused by the ohmic contacts, this time through their substantial bias dependency below 0.4 V, which considerably affects the magnitude of $s_{21}$.

We alternatively extract $v_p$ from the phase, $\theta$, of the measured $s_{21}$, whose absolute value vs. frequency is shown in Fig. 2(b) at various biases. Equation (1) yields the following expression for $\theta$ (as stated earlier, $4Z_0Z/(Z+2Z_0)^2$ of Eq. (1) has a negligible phase):

$$\theta = -\arctan\left[\frac{\sin(l_1\omega/v_p)+(A_2/A_1)\sin(l_2\omega/v_p)}{\cos(l_1\omega/v_p)+(A_2/A_1)\cos(l_2\omega/v_p)}\right], \quad (2)$$

which reduces to $\theta = \omega(l_1+l_2)/(2v_p)$ for $A_1 = A_2$, as expected. Equation (2) predicts that with a larger bias, thus with a larger $v_p$, $\theta$ increases more slowly with frequency, as evident in Fig. 2(b). We extract $v_p$ by fitting Eq. (2) to Fig. 2(b) at each gate bias. The $A_2/A_1$ ratios required for this fitting are extracted from the amplitudes of the dips and peaks of the $s_{21}$ magnitude in connection with Eq. (1); specifically, at a given bias, a total of five $A_2/A_1$ ratios are obtained over five frequency ranges, with each range spanning from a dip to its adjacent peak [APPENDIX C]; while the actual $A_2/A_1$ ratio must be constant at a given bias, the variations of the extracted $A_2/A_1$ values with frequencies at a given bias are caused by the frequency-dependent effects of the ohmic contacts and ungated 2DEG regions. Subsequently, fitting of Eq. (2) to Fig. 2(b) is performed in each of these five frequency ranges, yielding five values of $v_p$ at each bias. The ranges (blue bars) and medians (blue dots) of $v_p$ so extracted from the measured $s_{21}$ phase are shown in Fig. 3. As compared to $v_p$ extracted from the $s_{21}$ magnitude, this $s_{21}$-phase based extraction shows a greater immunity to the ohmic contact and ungated 2DEG effects: first, the frequency-dependent variation (sizes of the bars) is smaller; second, $v_p$ shows the expected monotonic increase with an increasing bias even below 0.4 V. At biases in excess of 0.4 V, the median values of $v_p$ extracted from the $s_{21}$ magnitude and those extracted from the $s_{21}$ phase show a match within 5 %. The high-fidelity $v_p$ extracted from $s_{21}$ phase ranges from ~$c$/330 (0.2 V) to ~$c$/250 (0.85 V), attesting to ultra-subwavelength confinement of the gated 2D plasmons.

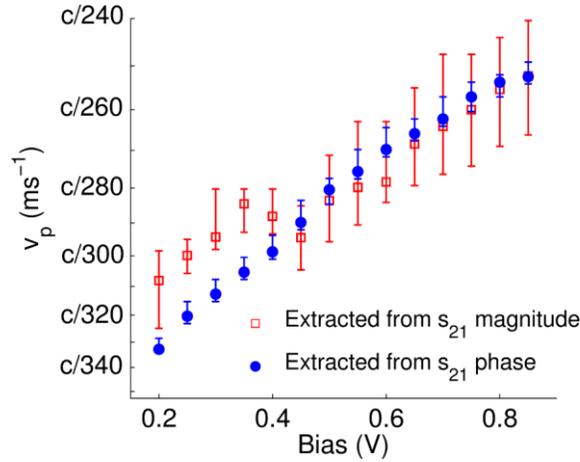

**Fig. 3:** The plasmonic velocities extracted from the measured $s_{21}$ magnitude (red squares and bars) and those extracted from the measured $s_{21}$ phase (blue dots and bars). The squares/dots and bars respectively represent the median values and frequency-dependent variations at given biases.

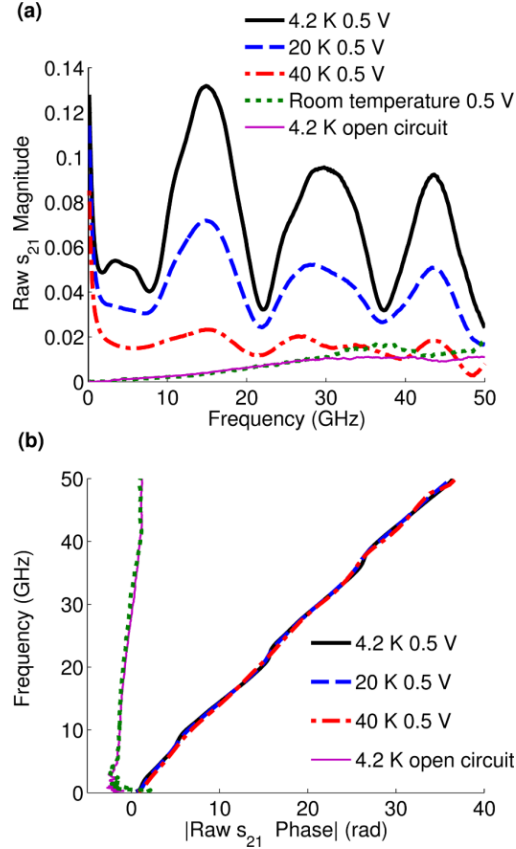

**Fig. 4:** (a) Magnitude and (b) phase of the interferometer's raw $s_{21}$ parameters at 4.2 K (black, solid), 20 K (blue, dashed), 40 K (red, semi-dashed), and room temperature (green, dotted), and of the actual open circuits' $s_{21}$ parameters at 4.2 K (magenta, thin).

We further confirm the plasmonic interferences by temperature-dependent measurements. As temperature rises, the electron scatterings are promoted to increase the ohmic resistance[15], degrading the 2DEG as a plasmonic medium. Quantitatively, while the plasmonic quality factor, $Q=\omega\tau$ ($\tau$: mean electron scattering time), is greater than unity at 4.2 K at GHz frequencies, $\tau$ and thus $Q$ decreases with temperature, making the observation of plasmonic waves and their interferences increasingly difficult. This is attested by temperature-dependent measurements of Fig. 4, where we show, for a reason that will be clear shortly, raw $s_{21}$-parameters without de-embedding the direct coupling between the two CPWs. As temperature rises towards 40 K, the $s_{21}$ magnitude gradually decreases, blurring the interference pattern [Fig. 4(a)]. At room temperature, the ohmic resistance becomes so high that the interferometer becomes effectively open-circuited; the non-zero raw $s_{21}$ magnitude at room temperature in Fig. 4(a) is due to the direct coupling between the CPWs. This is confirmed by the close similarity between this raw $s_{21}$ magnitude at room temperature and the $s_{21}$ magnitude of the actual open device at 4.2 K, where this open device is attained by biasing the gate at -0.4 V and depleting the 2DEG [APPENDIX B]. It is for this comparison that Fig. 4 presents raw *s*-parameters. The raw $s_{21}$ phase in Fig. 4(b) tells the same physics: up to 40 K, $s_{21}$ phase does not change appreciably, as the plasmonic wave, while increasingly masked by electron scatterings, observably maintains the same velocity. At room temperature, the plasmonic dynamics is completely masked, and the interferometer behaves as an open circuit.

Due to the ultra-subwavelength confinement of 2D plasmons ($v_p \sim c/300$), the interferometer achieves a high resolution. Since $\Delta\phi \equiv \omega(l_1-l_2)/v_p$ and $v_p \sim c/300$, a given path length difference, $l_1-l_2$, will lead to a $\sim 300/\sqrt{\kappa_{EM}}$ times larger phase difference in the plasmonic interferometer than in an electromagnetic wave interferometer whose wave speed is $c/\sqrt{\kappa_{EM}}$ ($\kappa_{EM}$ is the effective dielectric constant of the medium in the electromagnetic interferometer). So for the given $\Delta\phi$ resolution of the vector network analyzer, the plasmonic interferometer achieves a $\sim 300/\sqrt{\kappa_{EM}}$ times higher resolution than the electromagnetic interferometer at the same frequency. The plasmonic interferometer is also capable of a higher resolution detection of changes in the physical parameters on which $v_p$ depends, where this benefit is again attributed to the ultra-subwavelength confinement. For example, a change in the refractive index $\sqrt{\kappa}$ of the medium separating the gate and the 2DEG incurs a change in $\Delta\phi$ as $d(\Delta\phi)/d\sqrt{\kappa} = [\omega(l_1-l_2)/c] \times [c/(v_p\sqrt{\kappa})]$, while this derivative for an electromagnetic interferometer with a wave speed of $c/\sqrt{\kappa_{EM}}$ is[16] $d(\Delta\phi)/d\sqrt{\kappa_{EM}} = \omega(l_1-l_2)/c$. Thus the plasmonic interferometer is $\sim 300/\sqrt{\kappa}$ times more sensitive to the refractive index change, thus achieving a higher resolution by the same factor. The plasmonic interferometer can also detect the changes in the 2DEG electron density, *n*, and the effective electron mass, $m^*$, with high resolution, as $v_p$ depends on these quantities. They are quantities irrelevant to electromagnetic interferometers, thus, their detection is a bonus feature of our interferometer.

Despite the high-resolution advantage, the GHz proof-of-concept prototype presented in this paper is limited to cryogenic operation. At higher THz and infrared frequencies, 2D plasmons in GaAs/AlGaAs 2DEG and graphene can be observed at room temperature[12, 17], while maintaining the nature of subwavelength confinement. Scaling our design into this higher frequency regime may enable high-resolution THz and IR interferometric applications, such as biochemical detection, molecular spectroscopy, and precision modulation, at room temperature.

**ACKNOWLEDGEMENTS**

This work was supported by the Air Force Office of Scientific Research under contract number FA 9550-09-1-0369. Device fabrication was performed in part at the Center for Nanoscale Systems at Harvard University.

**APPENDIX A: Materials and fabrication**

Before device fabrication, the GaAs/AlGaAs 2DEG at 4.2 K has a carrier density, $n$, of $1.54 \times 10^{11}$ cm$^{-2}$ and a mobility, $\mu$, of $2.5 \times 10^6$ cm$^{-2}$/Vs in the dark. Under illumination, $n = 2.8 \times 10^{11}$ cm$^{-2}$ and $\mu = 3.9 \times 10^6$ cm$^{-2}$/Vs. At room temperature, $n = 3.76 \times 10^{11}$ cm$^{-2}$ and $\mu = 3.66 \times 10^3$ cm$^{-2}$/Vs in the dark. We etch the GaAs/AlGaAs sample by > 80 nm from the top surface to define the two 2DEG mesa paths of the interferometer (above the 2DEG, there is a 48 nm undoped Al$_{0.3}$Ga$_{0.7}$As layer, a 26 nm Si-doped Al$_{0.3}$Ga$_{0.7}$As layer, and a 6 nm GaAs cap). The ohmic contacts are created by thermally evaporating layers of Ni (5 nm), Au (20 nm), Ge (25 nm), Au (10 nm), Ni (5 nm), and Au (40 nm), followed by annealing at 420ºC for 50 seconds. The CPWs and the top gate are deposited by thermal evaporation of Cr (8 nm) and Au (500 nm).

We use the Sonnet electromagnetic field solver to design the 50-Ω CPWs[10]. The signal lines have a width of 24 μm and a length of 218 μm and are separated by 15 μm from the 123 μm-wide ground lines on each side. Each ohmic contact occupies an area of 6×24 μm$^2$. The separations between the top gate and the CPW signal lines on the left and right sides of the interferometer are 7 μm [Fig. 1]. These ungated 2DEG paths are far shorter than the gated 2DEG paths. This is to minimize the nonlinear dispersion effect of the ungated 2DEG regions.

**APPENDIX B: Calibration and de-embedding in *s*-parameter measurements**

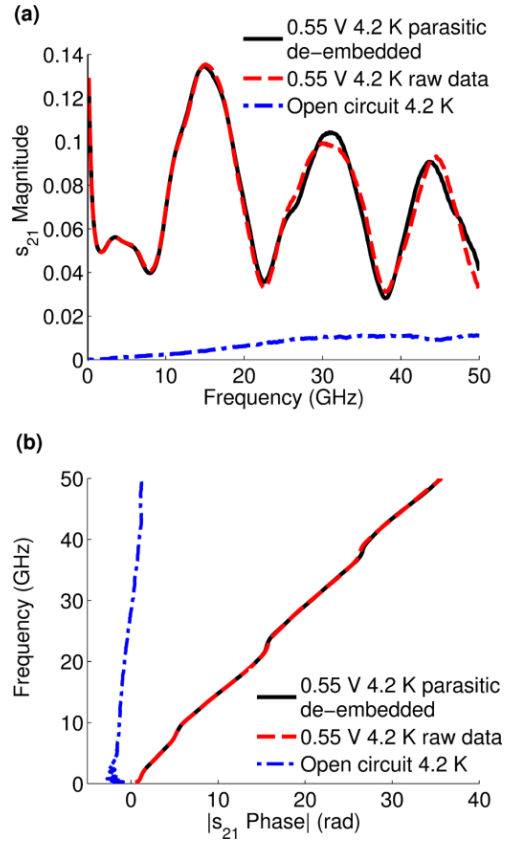

**Fig. A1:** (a) Magnitudes of raw $s_{21}$ (red, dashed), de-embedded $s_{21}$ (black, solid), and open-device $s_{21}$ at 4.2 K with a gate bias of 0.55 V. (b) Phases of the three $s_{21}$ parameters.

The measurements are done inside a Lakeshore cryogenic probe station. The Agilent E8364A vector network analyzer generates an *ac* signal up to 50 GHz with -45 dBm power reaching the device via ground-signal-ground microwave probes (100-μm pitch) and coaxial cables. The network analyzer, cables, and probes all have a characteristic impedance of 50 Ω. The network analyzer measures *s*-parameters. The delay and loss from the network analyzer up to the probe tips are calibrated out by using the NIST-style multi-line TRL calibration method[18]; this procedure involves measuring a set of *s*-parameters for CPWs of varying lengths fabricated on a separate GaAs substrate with no 2DEG present[10,11]. This calibration leads to the raw *s*-parameters, which include the effects of the interferometer, the phase delays through the on-chip CPWs, and the direct parasitic coupling between the two on-chip CPWs, which bypass the interferometer. The phase delays of the electromagnetic waves traveling through the CPWs are far smaller than the phase delays of the much slower plasmonic waves traveling through the interferometer, so we safely ignore the CPW phase delays. The effect of the parasitic coupling is separately measured by applying a gate bias of -0.4 V, thus depleting the 2DEG to imitate an open circuit [Fig. A1]. These open-device *s*-parameters are then de-embedded from the interferometer's raw *s*-parameters[19]. Figure A1 juxtaposes the raw and the de-embedded *s*-parameters at an example bias at 0.55 V. All the *s*-parameters discussed in the main text, except those in Fig. 4, are the de-embedded *s*-parameters.

## APPENDIX C: Extracted $A_2/A_1$ and $\alpha$

Figure A2(a) shows the $A_2/A_1$ ratios extracted from the $s_{21}$ magnitudes following the prescription given in the main text. Since $A_2/A_1 = e^{\alpha(l_1-l_2)}$, we can then extract the attenuation constant $\alpha$, [Fig. A2(b)]. The median values of $\alpha$ are around 8000 m$^{-1}$ at gate biases in excess of 0.4 V, where ohmic contact effects are less pronounced in the $s_{21}$ magnitudes.

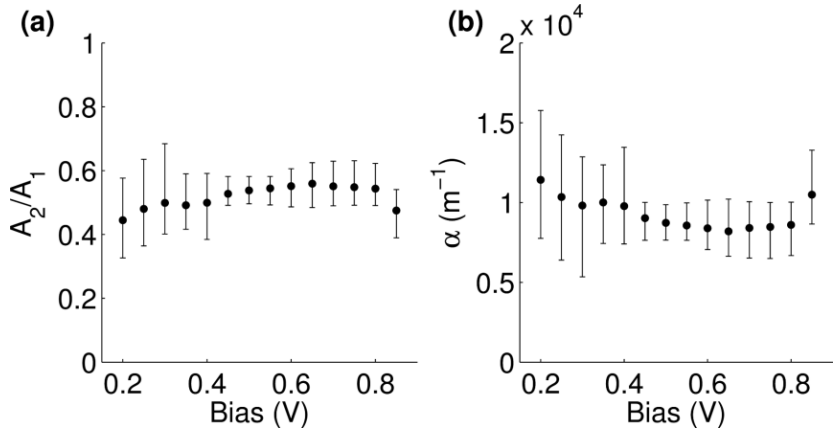

**Fig. A2:** (a) $A_2/A_1$ extracted from the $s_{21}$ magnitude. (b) $\alpha$ extracted from $A_2/A_1$ of part (a).

# APPENDIX D: On characteristic impedance, Z, of the plasmonic transmission lines and on the factor $4Z_0Z/(Z+2Z_0)^2$ of Eq. (1)

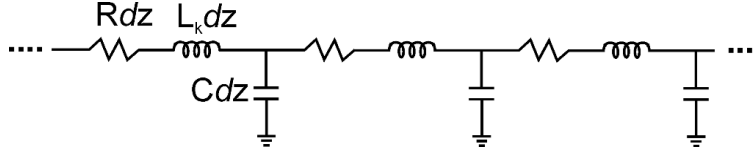

**Fig. A3:** The circuit model of the gated 2D plasmonic transmission line.

Both gated 2D plasmonic transmission lines of the interferometer have an identical 2DEG width $w$, thus the same characteristic impedance, $Z$. The gated 2D plasmonic line can be modeled as a distributed ladder network of kinetic inductors and capacitors[9, 10] [Fig. A3]; $L_k = m^*/ne^2w$ is the kinetic inductance per unit length[9-11], $C = \kappa\varepsilon_0 w/d$ is the capacitance per unit length, and $dz$ is an infinitesimal segment of the line. The ohmic resistance per unit length, $R = 1/(ne\mu w)$, in series with $L_k$ stems from the electron scatterings. The characteristic impedance, $Z$, of the plasmonic line is then given by $Z = \sqrt{(R+i\omega L_k)/(i\omega C)} = \sqrt{L_k/C}\sqrt{1-i/Q}$, where $Q = \omega L_k/R = \omega\tau$ is the quality factor. To evaluate $Z$, we should first know the values for $L_k$ and $R$ (on the other hand, $C$

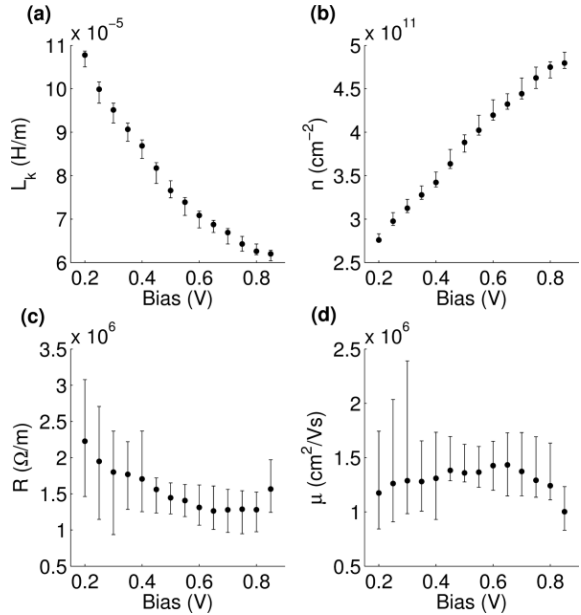

**Fig. A4:** (a) Extracted $L_k$. (b) Extracted $R$. (c) Extracted $n$. (d) Extracted $\mu$.

= $\kappa\varepsilon_0 w/d$ ~ $1.14\times10^{-8}$ F/m can be readily evaluated using the known geometric parameters). By using $v_p$ values extracted from the $s_{21}$ phase [Fig. 3] in $v_p = 1/\sqrt{(L_k C)}$, we extract $L_k$ values [Fig. A4(a)]. By noting that[20] $Q = \omega L_k/R$ on the one hand and $Q \sim \beta/2\alpha = \omega/(2v_p\alpha)$ on the other hand, we can write $R = 2\alpha v_p L_k$; then by using the extracted values of $\alpha$ [Fig. A2(b)], $v_p$ [Fig. 3], and $L_k$ [Fig. A4(a)] in this formula, we can extract $R$ [Fig. A4(b)]. With the extracted $R$ and $L_k$ values, we can evaluate $Z = \sqrt{(R+i\omega L_k)/(i\omega C)} = \sqrt{L_k/C}\sqrt{1-i/Q}$. As $Q = \omega L_k/R \sim 1$ occurs below 5 GHz, the imaginary part of $Z$ becomes increasingly small as frequency rises. Moreover, by substituting this $Z$ into $F \equiv 4Z_0 Z/(Z+2Z_0)^2$ of Eq. (1), we find that $F$ itself has a negligible imaginary part as compared to its almost constant real part over nearly all measurement frequency range [Fig. A5(a)]. That is, $F$ has a negligible phase [Fig. A5(b)] and a constant magnitude. This is self-consistent with our $v_p$ extraction from $s_{21}$ phase, where we ignored the phase of $F$.

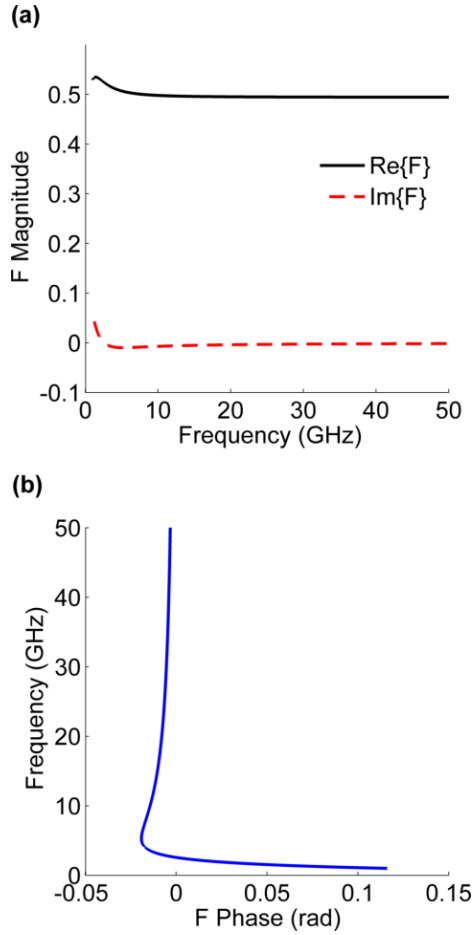

**Fig. A5:** (a) Real (black, solid) and imaginary (red, dashed) parts of $F$. (b) Phase of $F$. Bias: 0.55 V.

The extracted $L_k$ and $R$ values also allow us to extract the electron density, $n$, and the mobility, $\mu$ [Fig. A4(c) and Fig. A4(d)]. The median values of the extracted $\mu$ lie between $1 \times 10^6 \sim 1.5 \times 10^6$ cm$^{-2}$/Vs, which are comparable to, but justifiably smaller due to fabrication steps than, the mobility of the pristine sample ($2.5 \times 10^6$ cm$^{-2}$/Vs).

**APPENDIX E: Derivation of Eq. (1)**

Figure A6(a) illustrates our interferometer consisting of two plasmonic transmission lines '1' and '2' along with the two on-chip CPWs 'A' and 'B'. When an electromagnetic wave is launched onto CPW A, multiple transmissions and reflections will occur at the two junctions in the figure, and multiple waves will appear at CPW B through many different signal pathways. Superposition of these multiple waves represents the total transmitted wave. There are two 1$^{st}$-order signal pathways from CPW A to CPW B that exhibit the lowest degree of loss: A→1→B and A→2→B. There are eight 2$^{nd}$-order signal pathways from CPW A to CPW B that exhibit the second lowest degree of loss (in what follows, a path number with no prime signifies left-to-right propagation on that path, and a path number with prime signifies right-to-left propagation): A→1→1'→1→B, A→1→1'→2→B, A→1→2'→1→B, A→1→2'→2→B, A→2→1'→1→B, A→2→1'→2→B, A→2→2'→1→B, and A→2→2'→2→B. Similarly we can identify higher-order signal pathways.

We first calculate the contribution of the 1$^{st}$-order signal pathways (A→1→B and A→2→B) to $s_{21}$. Since $s$-parameters are defined in terms of power waves[21], we start by calculating *local* transmission coefficients for A→1, 1→B, A→2, and 2→B for power waves. Figure A6(b) shows the left junction. The incoming power wave $a_A^+$ on CPW A produces, at the junction, transmitted power wave $a_1^+$ and $a_2^+$ on path 1 and 2 (as well as the reflected power wave $a_A^-$, which is not involved in the $s_{21}$ calculation with the 1$^{st}$-order signal pathways), where[21]:

$$a_A^+ = \frac{V_1^+ + Z_0(I_1^+ + I_2^+)}{2\sqrt{Z_0}}; \quad a_1^+ = \frac{V_1^+ + ZI_1^+}{2\sqrt{\text{Re}\{Z\}}}; \quad a_2^+ = \frac{V_2^+ + ZI_2^+}{2\sqrt{\text{Re}\{Z\}}}. \tag{A1}$$

Here $V_1^+$ and $V_2^+$ ($=V_1^+$ due to the shunt connections of path 1 and path 2 at the junction) are the voltages of the plasmonic waves transmitted onto path 1 and path 2, read at the junction; $I_1^+$ and $I_2^+$ ($=I_1^+$ because path 1 and path 2 have the same characteristic impedance, $Z$) are the currents of the plasmonic waves transmitted onto path 1 and path 2, read at the junction. Then the *local* transmission coefficients[20] $t_{A\to 1} = a_1^+/a_A^+$ and $t_{A\to 2} = a_2^+/a_A^+$ are given by

$$t_{A\to 1} = t_{A\to 2} = \sqrt{\frac{Z_0}{\text{Re}\{Z\}}} \frac{2Z}{Z + 2Z_0}. \tag{A2}$$

Similarly, we can calculate the *local* transmission coefficients $t_{1\to B}$ and $t_{2\to B}$ as:

$$t_{1\to B} = t_{2\to B} = \sqrt{\frac{\text{Re}\{Z\}}{Z_0}} \frac{2Z_0}{Z + 2Z_0}. \tag{A2}$$

Then the overall A-to-B transmission, $s_{21}$, through the 1$^{\text{st}}$-order signal pathways is given by

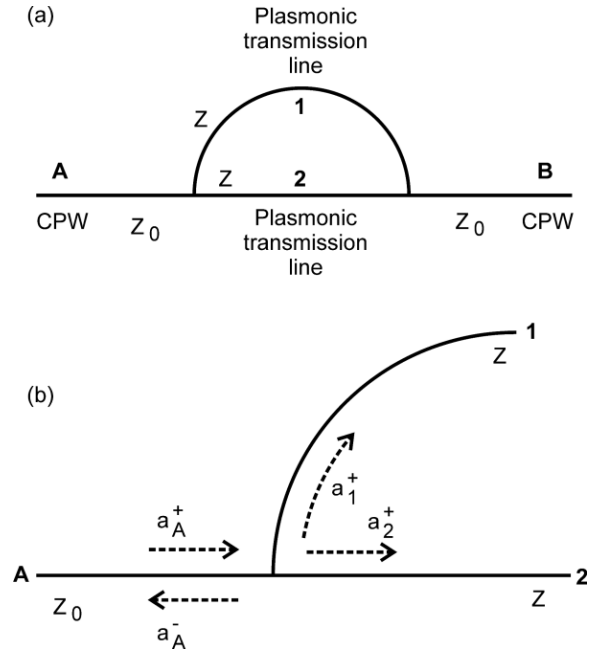

**Fig. A6** (a) Schematic showing the interferometer's two plasmonic paths (1 and 2) along with the two on-chip CPWs (A and B). (b) Illustration of local transmissions and reflections for the on-coming power wave from CPW A.

$t_{A \to 1} t_{1 \to B} e^{-\alpha l_1} e^{-i\beta l_1} + t_{A \to 2} t_{2 \to B} e^{-\alpha l_2} e^{-i\beta l_2}$, that is:

$$s_{21} = \frac{4ZZ_0}{(Z+2Z_0)^2} \left[ e^{-\alpha l_1} e^{-i\beta l_1} + e^{-\alpha l_2} e^{-i\beta l_2} \right] \quad [1^{st} \text{ order}]. \tag{A3}$$

We now consider the contribution of the $2^{nd}$-order signal pathways to $s_{21}$. Given that $\alpha \sim 8000$ m$^{-1}$ (APPENDIX C), traversing path 1 ($l_1 \sim 191$ μm) once will reduce the amplitude by a factor of $\exp(-\alpha l_1) \sim 0.22$ and traversing path 2 ($l_2 \sim 120$ μm) once will reduce the amplitude by a factor of $\exp(-\alpha l_2) \sim 0.38$. Since each of the eight $2^{nd}$-order signal pathways involves traversing path 1 and/or path 2 a total of three times, each pathway will suffer significant attenuation. In addition, each $2^{nd}$-order pathway involves two additional local reflections and/or transmissions, which further attenuates the signal. The eight substantially attenuated signals superpose in CPW B, but they have generally different phases, thus, their superposition does not help much in countering the attenuation. All in all, the contribution from the $2^{nd}$-order pathways to $s_{21}$ is negligibly small, which is confirmed by the actual calculation of the $2^{nd}$-order contribution [Fig. A7]. In sum, it is sufficient to calculate $s_{21}$ up to the $1^{st}$ order as in Eq. (A3), which is Eq. (1) of the main text.

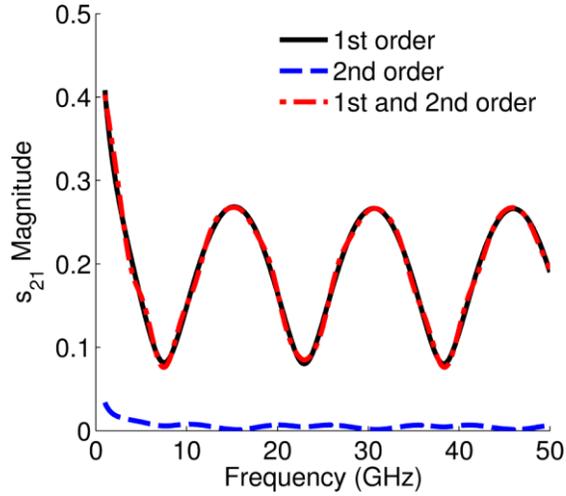

**Fig. A7:** Contributions of the $1^{st}$ (black, solid) and $2^{nd}$ (blue, dashed) order signal pathways to the $s_{21}$ magnitude and their sum (red, semi-dashed). Physical parameters used ($L_k$, $R$, $\alpha$ and $\beta$) for this plot are those (median values) extracted at 0.55 V.

**REFERENCES**

[1] H. F. Schouten, N. Kuzmin, G. Dubois, T. D. Visser, G. Gbur, P. F. A. Alkemade, H. Blok, G. W. 't Hooft, D. Lenstra, and E. R. Eliel, *Phys. Rev. Lett.* **94**, 053901 (2005).

[2] R. Zia, and M. L. Brongersma, *Nature Nanotechnology* **2**, 426 (2007).

[3] S. I. Bozhevolnyi, V. S. Volkov, E. Devaux, J.-Y. Laluet and T. W. Ebbesen, *Nature* **440**, 508 (2006).

[4] J. Feng, V. S. Siu, A. Roelke, V. Mehta, S. Y. Rhieu, G. T. R. Palmore and D. Pacifici, *Nano Letters* **12**, 602 (2012).

[5] W. L. Barnes, A. Dereux and T. W. Ebbesen, *Nature* **424**, 824 (2003).

[6] F. Stern, *Phys. Rev. Lett.* **18**, 546 (1967).

[7] S. J. Allen, Jr., D. C. Tsui and R. A. Logan, *Phys. Rev. Lett*. **38**, 980 (1977).

[8] I. V. Kukushkin, J. H. Smet, S. A. Mikhailov, D. V. Kulakovskii, K. von Klitzing and W. Wegscheider, *Phys. Rev. Lett*. **90**, 156801 (2003).

[9] P. J. Burke, I. B. Spielman, J. P. Eisenstein, L. N. Pfeiffer and K. W. West, *App. Phy. Lett.* **76**, 745 (2000).

[10] W. F. Andress, H. Yoon, K. Y. M. Yeung, L. Qin, K. West, L. Pfeiffer and D. Ham, *Nano Letters*, **12**, 2272 (2012).

[11] H. Yoon, K. Y. M. Yeung, V. Umansky, D. Ham, *Nature*, **488**, 65 (2012)

[12] L. Ju, B. Geng, J. Horng, C. Girit, M. Martin, Z. Hao, H. A. Bechtel, Z. Liang, A. Zettl, Y. R. Shen and F. Wang, *Nature Nanotech*. **6**, 630 (2011)

[13] W. F. Andress and D. Ham, *IEEE J. Solid-State Circuits*, **40**, 638 (2005)

[14] A. Eguiluz, T. K Lee, J. J. Quinn and K. W. Chiu, *Phys. Rev. B*, **11**, 4989 (1975).

[15] B. J. F.Lin, D. C. Tsui, M. A. Paalanen and A. C. Gossard, *Appl. Phys. Lett*. **45**. 695 (1984)

[16] R. Ince and R. Narayanaswamy, *Analy. Chim. Acta*, **569**, 1 (2006).

[17] Y. M. Meziani, H. Handa, W. Knap, T. Otsuji, E. Sano, V.V. Popov, G. M. Tsymbalov, D. Coquillat and F. Teppe. *Appl. Phys. Lett.* **92**, 201108 (2008).

[18] R. B. Marks, *IEEE Transactions on Microwave Theory and Techniques*, **39**, 1205 (1991).

[19] A. Aktas and M. Ismail, *IEEE Circuits Devices Mag*. **17**, 8 (2001).

[20] D. M. Pozar, *Microwave Engineering 3$^{rd}$ Edition* (John Wiley & Sons, 2005).

[21] K. Kurokawa, *IEEE Transactions on Microwave Theory and Techniques*, **13**, 194 (1965).